\documentclass{ws-ijmpa}

\begin{document}

\markboth{Sinya Aoki}{QCD Phases in Lattice QCD}

\catchline{}{}{}{}{}

\title{QCD Phases in Lattice QCD}

\author{Sinya AOKI}
\address{Graduate School of Pure and Applied Sciences, University of Tsukuba,
Tsukuba, \\ Ibaraki 305-8571, Japan, email: saoki@het.ph.tsukuba.ac.jp  and \\
Riken BNL Research Center, Brookhaven National Laboratory, Upton, NY11973, USA
}


\maketitle


\begin{abstract}
I review the recent status of lattice QCD calculations at non-zero
density.
\end{abstract}



\section{Introduction}
It is believed that QCD undergoes phase transitions
as a temperature $T$ and a chemical potential $\mu$ increase.
For example, an expected phase structure is found in Fig.1 of 
ref.~\refcite{hands}.
To theoretically investigate what really happens at non-zero $T$ and $\mu$
in QCD, lattice calculations seem ideal. As reviewed 
in ref.~\refcite{ejiri1}, lattice QCD investigations are indeed 
very successful at finite $T$ but $\mu=0$. 
On the other hand,
lattice QCD simulations at non-zero $\mu$ have remained to be
extremely difficult for a long time, since a complex nature of the lattice QCD
action makes a direct application of
straightforward Monte-Carlo methods impossible.
Recently, however, several new techniques have been proposed to overcome
this difficulty.
In this talk I mainly review recent progresses in lattice QCD
at non-zero $\mu$.

\section{Lattice QCD at non-zero $\mu$}
A possible solution to the complex action problem at non-zero $\mu$
was proposed in ref.~\refcite{glasgow}, where a reweighting method
from $\mu=0$ was employed. This method, called the Glasgow method, 
however failed since an overlap between gauge ensembles at 
$\mu=0$ and  at $\mu\not=0$ is exponentially small in the volume.

Recently two new methods have been proposed for small $\mu$ near the
critical temperature $T_c$.
One is a variant of the Glasgow method\footnote{For a reweighting of only
$\mu$ by the Taylor expansion, see ref.~\refcite{GG}.
}
: Instead of reweighting only
$\mu$ from $\mu=0$, a multi-parameter reweighting in both $\mu$ and $T$
from $\mu=0$ and $T=T_c(\mu=0)$ 
is employed\cite{FK1,FK2,FK3,B-S}.
The other method has employed an imaginary chemical potential $\mu= i\mu_I$
to avoid the complex action problem\cite{DEL,FP}. 
Calculations in both methods so far are restricted to $N_t = 4$, where the
temperature is given by $ T =1/(N_t a)$.

\subsection{Multi-parameter reweighting method}
I first discuss the multi-parameter reweighting method, in which the
QCD partition function at non-zero $\mu$ is rewritten as
\begin{eqnarray}
Z(\beta, \mu, m) &=&\int {\cal D}U \exp\left[ - S_G(\beta, U)\right]
\det D(U,\mu,m) \\
&=&
\int {\cal D}U  \exp\left[ - S_G(\beta^\prime, U)\right]
\det D(U,\mu^\prime,m) \nonumber \\
&\times& \left\{
\exp\left[-S_G(\beta,U)+S_G(\beta^\prime,U)
\right]
\frac{\det D(U,\mu,m)}{\det D(U,\mu^\prime m)}
\right\}
\end{eqnarray}
where $U$ represents a gauge configuration, $\beta = 6/g^2$, 
$m$ is a quark mass and $\mu$ is a (quark) chemical potential. 
Here $S_G$ is a gauge action and $\det D$ is a 
quark determinant. 
In the actual calculation, gauge configurations are generated at
$\mu^\prime = 0$ and $\beta^\prime \simeq \beta_c(m,0)$,
where $\beta_c (m,\mu)$ is a critical coupling at 
$\mu$ and $m$.
Since both $\beta$ and $\mu$ are 
used in the reweighting, this method is called a multi-parameter reweighting. 
Once configurations are generated at $\mu^\prime$ and $\beta^\prime$, 
one can treat the term in $\{\cdots \}$ as a part of observables.

Fodor and Katz\cite{FK3} have employed the multi-parameter 
reweighting method
starting from gauge configurations generated by
2+1 flavor KS quark action with the plaquette gauge action 
at $\mu^\prime=0$ and $\beta^\prime = \beta_c$.
Since they have exactly evaluated $\det D(U,\mu,m)$ appeared in the 
reweighting factor, lattice volumes in their calculations 
are restricted to $L_s^3\times 4$ with $L_s = 6,8,10, 12$.
Dynamical quark masses in the simulation are 
$m_{ud}/T = 0.0368$ ($ m_\pi/m_\rho = 0.188(1)$) 
and
$
m_s/T = 1.0$ ($ m_\pi/m_K = 0.267(1)$),
which are very close to experimental values, $m_\pi/m_\rho = 0.179$ and
$m_\pi/m_K = 0.267$.

\begin{figure}
\centerline{\psfig{file=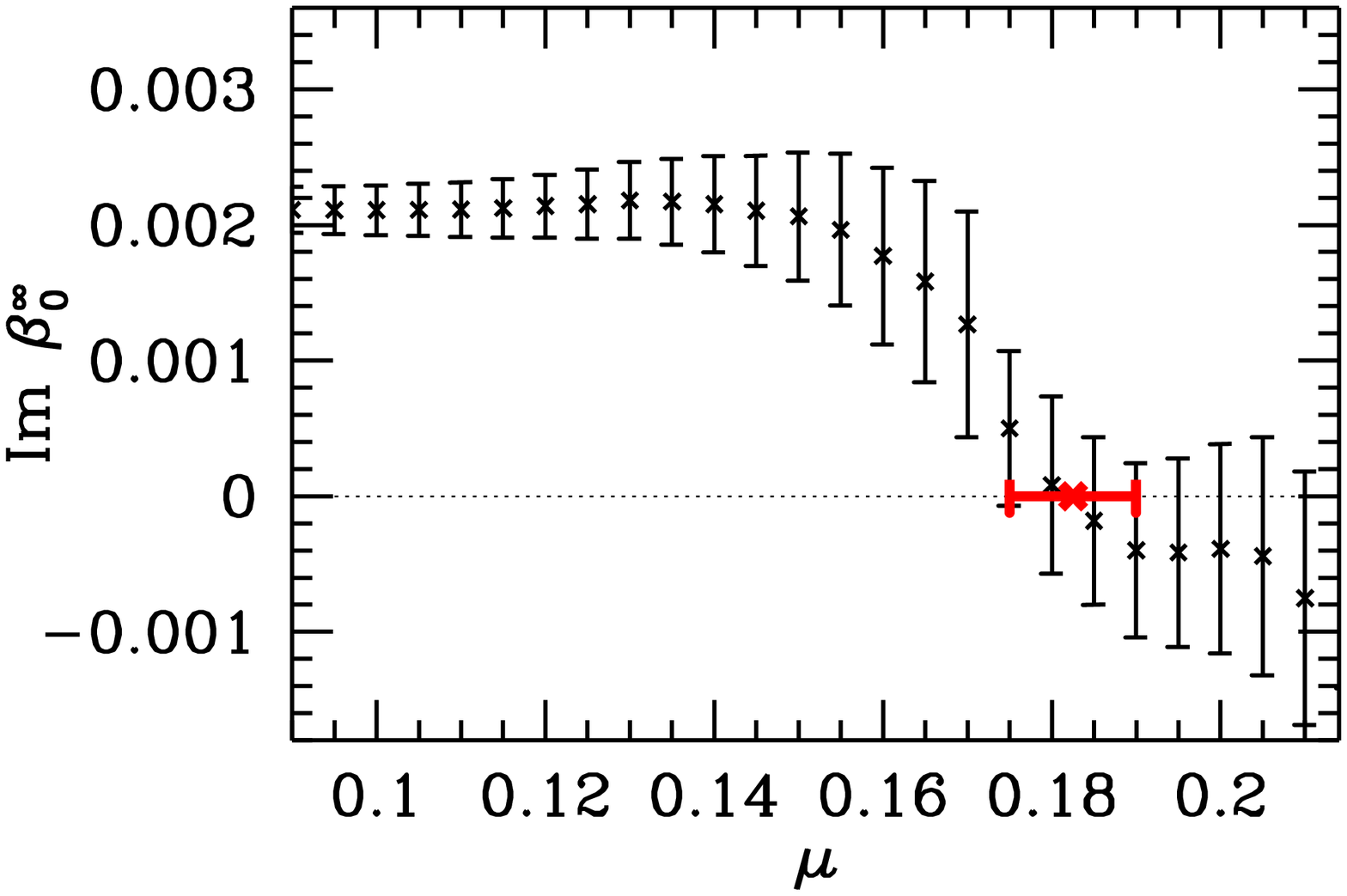,width=6.5cm}
\psfig{file=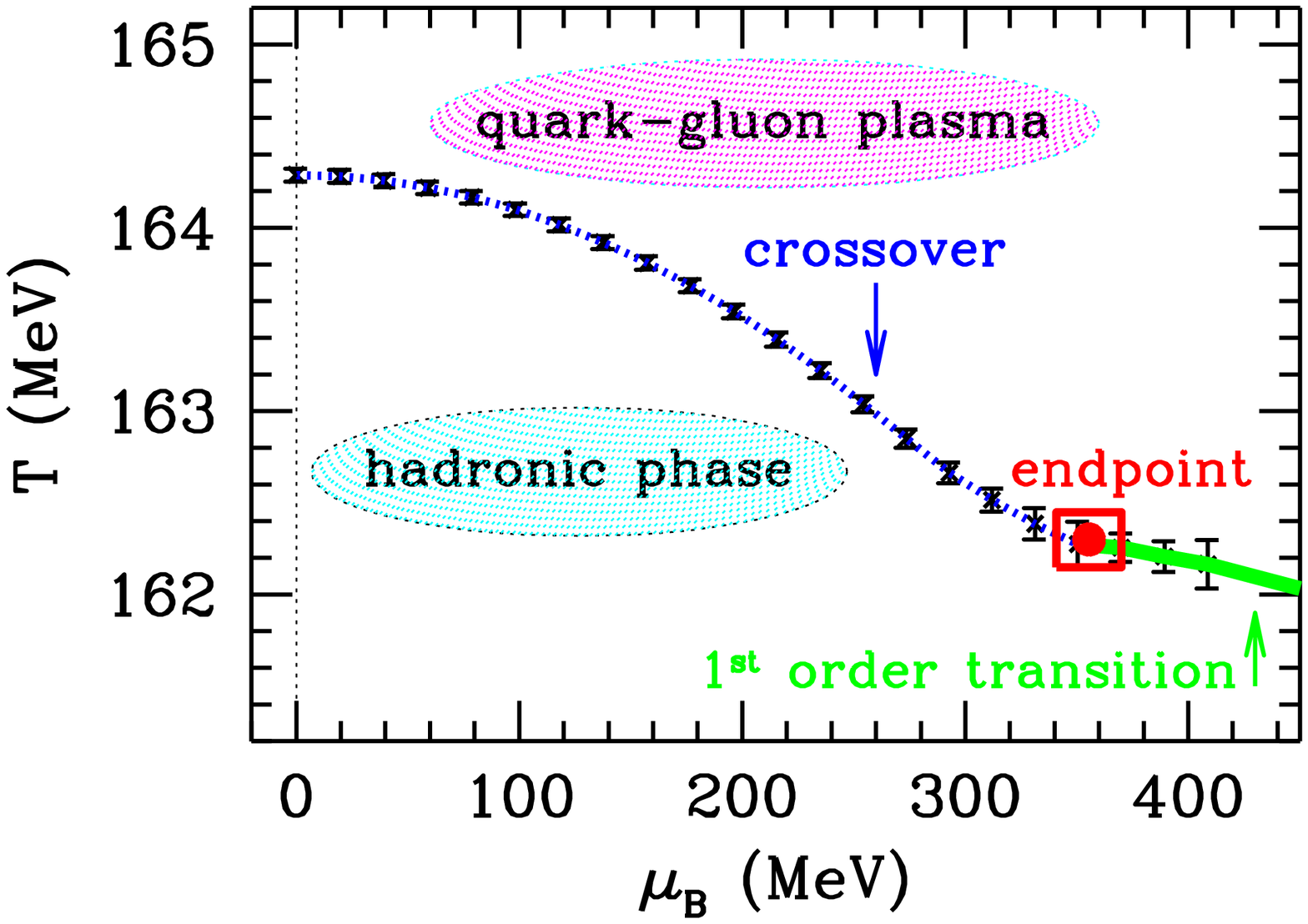,width=6.5cm}}
\caption{(Left) Im $\beta_0^\infty$ a s a function of the chemical 
potential. (Right) The phase diagram in physical unit.}
\label{fig:FK}
\end{figure}

In order to distinguish the first order phase transition from the crossover,
they investigated a volume dependence of Lee-Yang zeros\cite{LY} 
$\beta_0$, which are zeros of the partition function $Z(\beta,\mu,m)$ 
in the complex $\beta$ plane. Since there is no real phase transition in the 
finite volume, ${\rm Im}\, \beta_0 \not=0$ at $V=L_s^3 \not= \infty$.
If $\lim_{V\rightarrow\infty}{\rm Im}\, \beta_0 =0$, there exists
the first order phase transition at $\beta = \beta_0$,
while if $\lim_{V\rightarrow\infty}{\rm Im}\, \beta_0 \not= 0$,
only a crossover appears at $\beta = {\rm Re}\, \beta_0$.
In the left of Fig.\ref{fig:FK}, taken from ref.~\refcite{FK3},
${\rm Im}\, \beta_0^\infty$, obtained by extrapolating $\beta_0$ to the 
infinite volume as $\beta_0(V) = \beta_0^\infty + c/V$, 
is shown as a function of $\mu a$. 
The critical endpoint, which separate the first order phase transition line 
from the crossover line, is estimated to be $\mu a\simeq 0.18$.
Converting $({\rm Re}\, \beta_0^\infty, \mu a)$ to the physical unit
$(T_c, \mu_B=3\mu)$, where $\mu_B$ is a baryonic chemical potential,
a phase diagram for QCD at non-zero $T$ and $\mu$ is given
in the right of Fig.\ref{fig:FK}, also taken from ref.~\refcite{FK3},
and the critical endpoint becomes $(T_E, \mu_B^E)=(162(2), 360(40))$
MeV\cite{FK3}.

Instead of evaluating $\det D(U,\mu,m)$ exactly, Bielefeld-Swansea group
has calculated it by the Taylor expansion,
\begin{eqnarray}
\ln \left[
\frac{\det D(U,\mu,m)}{\det D(U,0,m)}
\right] = R_1\mu + R_2\mu^2 + \cdots ,
\end{eqnarray}
to perform the simulation on a relatively large volume, $16^3\times 4$.
Gauge configurations are generated by 
the improved gauge action and
the $N_f=2$ p4-improved KS quark action at $m/T =0.4, 0.8$,
corresponded to $m_\pi/m_\rho \simeq 0.7, 0.85$.
The transition temperature was estimated from the peak position of 
susceptibilities $\chi_{\bar\psi\psi}$ and $\chi_L$, where $L$ is the
Polyakov loop. The result of the critical temperature $T_c(\mu)$
as function of the quark chemical potential $\mu_q$ can be found in 
Fig.16 of ref.~\refcite{B-S}. Errors of the reweighting remain small at 
$\displaystyle\frac{\mu_q}{T} \le 0.4$, but 
they increase as $\mu_q$ becomes larger.

\subsection{Imaginary chemical potential}
Since the quark determinant $\det (U, i\mu_I, m)$ becomes real
for the imaginary chemical potential $i \mu_I$,
direct Monte-Carlo simulations are possible in this case\cite{DEL,FP}.
One then fits the critical temperature obtained at the imaginary chemical 
potential in the form of the Taylor expansion:
\begin{eqnarray}
T_c(i\mu_I) = T_c(0) + c_2 \mu_I^2 + O(\mu_I^4),
\end{eqnarray}
as seen in Fig. 9 of ref.~\refcite{FP} for $N_f=2$
and in Fig.6 of ref.~\refcite{DEL} for $N_f=4$.
Assuming an analyticity at $\mu=0$, one immediately obtain 
the critical temperature at real $\mu$:
\begin{eqnarray}
T_c(\mu) &=& T_c(0) - c_2 \mu^2 + O(\mu^4).
\end{eqnarray}
In this method,
a restriction that 
$\vert \bar\mu_I\equiv \frac{\mu_I}{T}\vert \le \frac{\pi}{3}$
exists, due to periodicity and symmetry, which leads to
\begin{eqnarray}
Z(\bar\mu_I = \frac{\pi}{3}+x) &=& Z(\bar\mu_I = \frac{\pi}{3} - x),
\end{eqnarray}
as explicitly confirmed in Fig.6 of ref.~\refcite{FP}.

\subsection{Comparison of various results}
The critical temperature $T_c$ can be parameterized as a function of 
the baryonic chemical potential $\mu_B$ as
\begin{eqnarray}
\frac{T_c(\mu_B)}{T_c(0)}= 1 - C \frac{\mu_B^2}{T_c(0)^2} .
\end{eqnarray}
Results for $C$ from various simulations are accumulated in 
table~\ref{tab:AllC}, which indicates that $C$ becomes larger as $N_f$ 
increases. Further investigations including the quark mass dependence,
however, will be required for the definite conclusion on the $N_f$
dependence of $C$.

A critical endpoint $(\mu_B^E, T_E)$ is a point on the 
critical temperature line which separates the cross-over from the first order 
phase transition.
Results of the critical endpoint from two groups are given in 
table~\ref{tab:CEP}, from which one may notice that the location of
the critical endpoint strongly depends on the quark mass.
Again further investigations will be required for the reliable estimate
on the location of the critical endpoint for the physical case.
\begin{table}[bt]
\tbl{Values of $C$ from various simulations. Here Imag, MulParRew, MulParRewTay
represent the imaginary chemical potential, the multi-parameter reweighting and
the multi-parameter reweighting with the Taylor expansion, respectively.
}
{\begin{tabular}{llllll} \toprule
$N_f$ & C & $m_{ud}/T$ & $m_{s}/T$  & quark & method$^{\rm ref.}$ \\
\colrule
2 & 0.0056(4) & 0.1 & - & KS & Imag\cite{FP}  \\
  & 0.008(3)  &0.4,0.8 & - & p4 & MulParRewTay\cite{B-S,B-S2} \\
\colrule
2+1 & 0.0032 & 0.0368 & 1.0 & KS & MulParRew\cite{FK3} \\
\colrule
3 & 0.0028(7) &\multicolumn{2}{c}{0.4}  & p4 &  MulParRewTay\cite{B-S2} \\
  & 0.013(5)  &\multicolumn{2}{c}{0.02} & p4 &  MulParRewTay\cite{B-S2} \\
  & 0.00678(10) &\multicolumn{2}{c}{0.1-0.16}& KS  & Imag\cite{FP2}($\mu^4$
included) \\
\colrule
4 & 0.011 & \multicolumn{2}{c}{0.2} & KS & Imag\cite{DEL} \\
\botrule
\end{tabular}}
\label{tab:AllC}
\end{table}

One may convert the mass dependence of the critical endpoint to
the $\mu$ dependence of the critical quark mass as
\begin{eqnarray}
\frac{m_c(\mu_q)}{m_c(0)} &=& 1 + D \left(\frac{\mu_q}{\pi T_c}\right)^2 ,
\end{eqnarray}
where the critical quark mass separates the first order phase transition from 
the cross-over: the phase transition is the first order at $m < m_c$
while it becomes cross-over at $ m > m_c$. 
For the $N_f=3$ case,
The imaginary chemical potential method\cite{FP2},
using the KS quark action,
gives $D=0.84(36)$ and $m_c(0)/T=0.123(1)$,
while the multi-parameter reweighting method with the Taylor 
expansion\cite{B-S2}, using the p4-improved KS quark action, 
leads to $ D=690$ and $m_c(0)/T=0.0028(16)$.
Although errors are large and quark actions are different,
a discrepancy between two results is huge.
We will have to understand and resolve this discrepancy, in order to make
the definite conclusion for the phase structure at non-zero
chemical potential.

\begin{table}[bt]
\tbl{The critical endpoints $(\mu_B^E, T_E)$,
together with the critical temperature $T_c$ at $\mu_B = 0$.}
{\begin{tabular}{llllllll} \toprule
$N_f$ & $\mu_B^E$ MeV & $T_E$ MeV& $ T_c$ MeV & $m_{ud}/T$ & $m_{s}/T$  
& quark & method$^{\rm ref.}$ \\
\colrule
2+1 & 725(35) & 160(4)& 172(3) & 0.1 & 0.8 & KS & Lee-Yang zero\cite{FK2}  \\
  & 360(40) & 162(2)& 164(2) & 0.0368 & 1.0 & KS & Lee-Yang zero\cite{FK3}  \\
\colrule
3 & 0 & - & -  & 0.0028(16) & 0.0028(16) & p4 &  Binder cumulants\cite{B-S2}\\
  & 156(30) & - & - & 0.02 & 0.02 & p4 &   Binder cumulants\cite{B-S2}\\ 
\botrule
\end{tabular}}
\label{tab:CEP}
\end{table}

\section{Some remarks}
\subsection{Reweighting vs. imaginary chemical potential}
\begin{figure}[b]
\centerline{\psfig{file=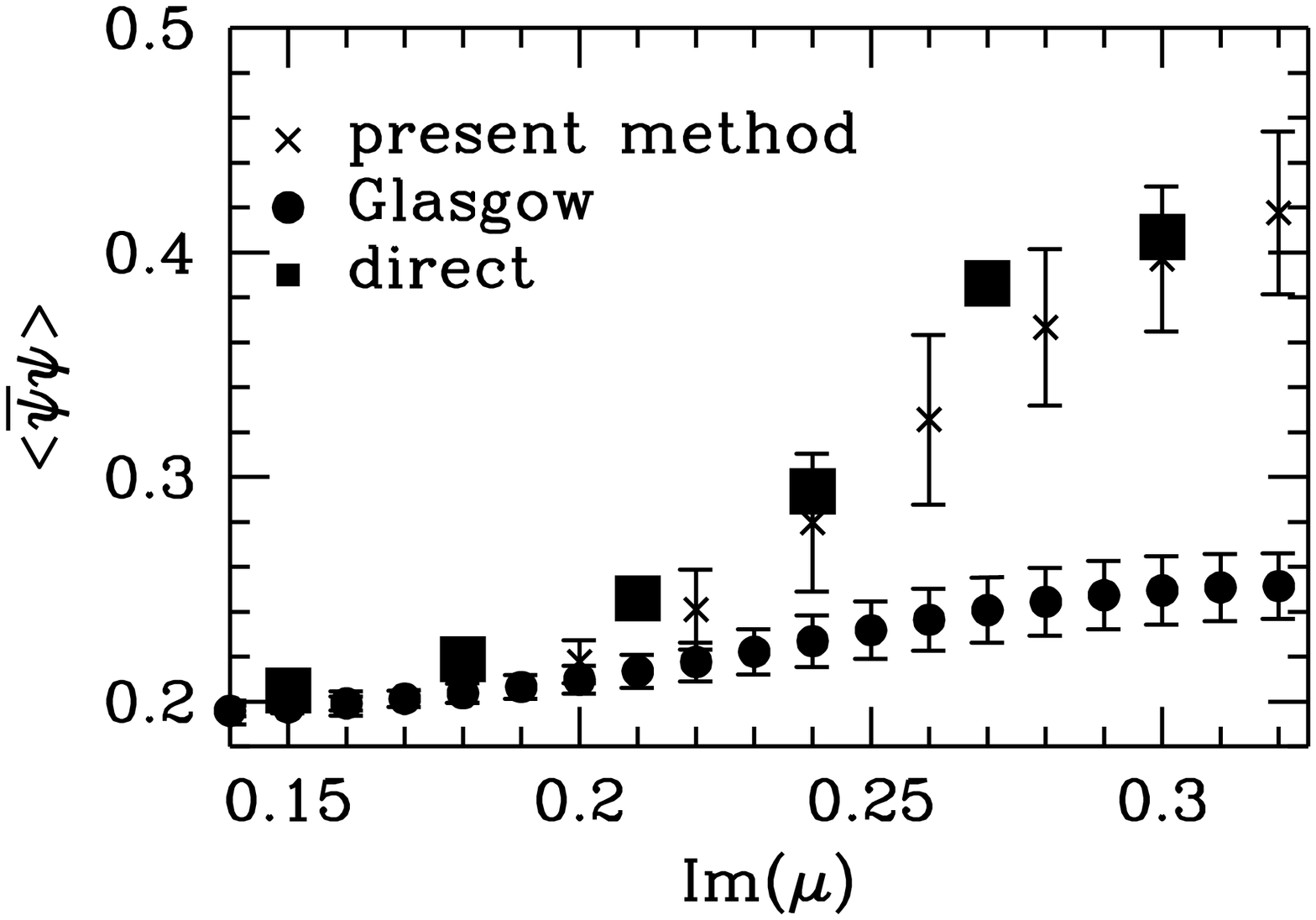,width=6.5cm}
\psfig{file=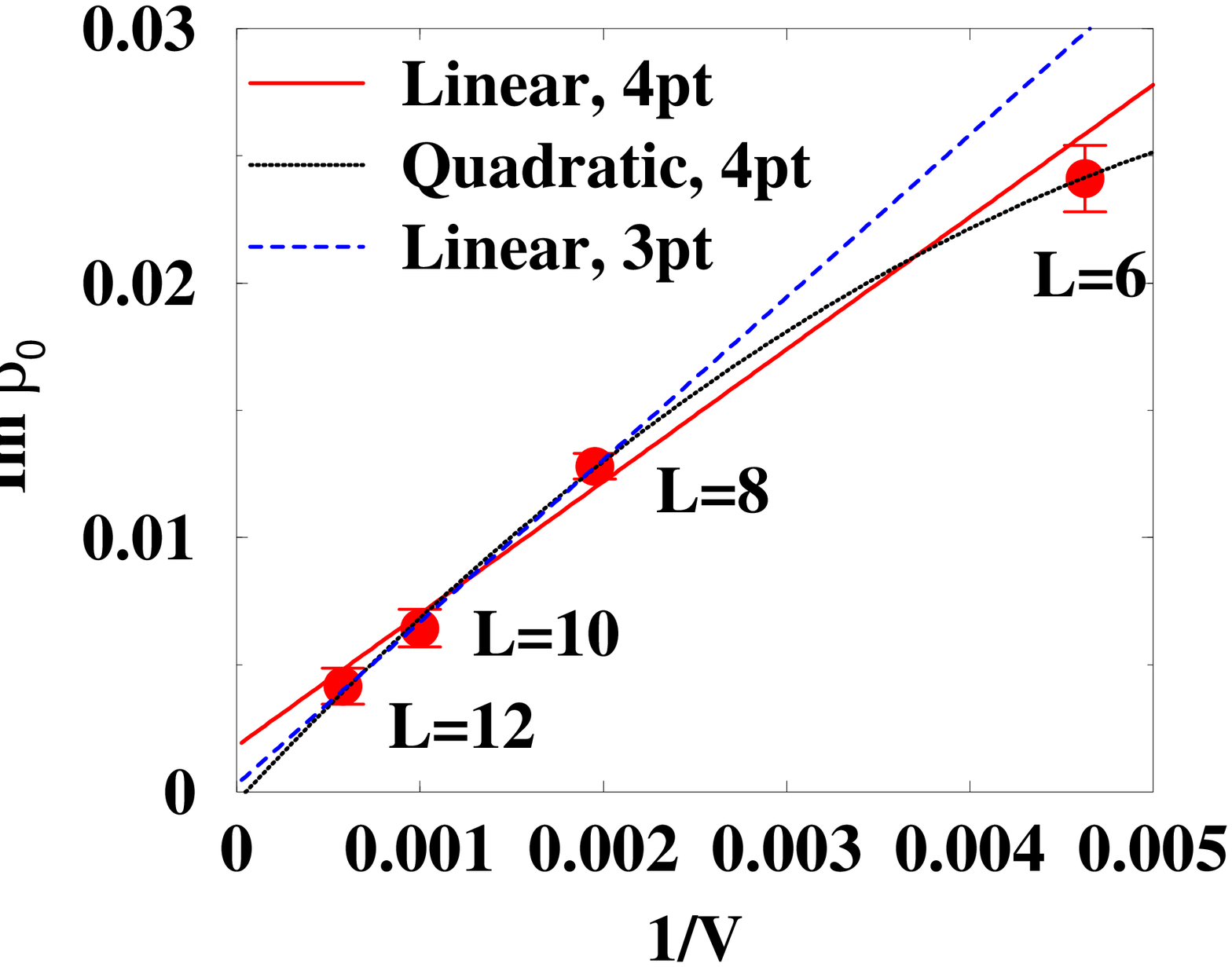,width=5.5cm}}
\caption{(Left) $\langle \bar\psi \psi\rangle$ as a function of
Im $\mu$ for the direct simulation(squares), the multi-parameter reweighting
(crosses) and the Glasgow method(circles).
(Right) Im $\beta_0$ as a function of $1/V$ at $\mu a = 0.16$, 
together with the linear 
fit(solid line), the quadratic fit(dotted line) and the linear fit for 
larger 3 volumes(dashed line).}
\label{fig:compare}
\end{figure}

One can check a reliability of the reweighting method by comparing it
with the direct calculation in the case of the imaginary chemical potential. 
In the left of Fig.~\ref{fig:compare}, 
taken from ref.~\refcite{FK1},
$\langle \bar \psi \psi \rangle$ is plotted as a function of $\mu_I$.
The multi-parameter reweighting method agrees with the direct calculation,
while the Glasgow method does not.
This comparison may suggest that the multi-parameter reweighting is
more reliable. As mentioned in the previous subsection, however,
the result must be symmetric in $\mu_I$ at $\mu_I = \pi/12 \simeq 0.26 $.
This symmetric property is clearly violated in both mult-parameter reweighting
method and the direct calculation.
It is likely that the violation may be caused by the insufficient statistics 
for both methods. Therefore, in the case of the reweighting
method, one should always check whether the symmetric property is
satisfied or not for the imaginary chemical potential, using the same 
ensemble employed for the reweighting of real $\mu$. 

\subsection{Cautions to the Lee-Yang zero analysis}
It has been pointed out that the partition function 
becomes zero in the infinite volume limit for all $\mu\not= 0$:
$
\displaystyle\lim_{V\rightarrow\infty} Z(\mu \not= 0, V) = 0
$,
due to the sign problem of the complex action\cite{ejiri2}.
Therefore one should check the volume dependence of
not only the 1st Lee-Yang zero but also the 2nd, 3rd, $\cdots$ Lee-Yang zeros,
in order to distinguish the first order phase transition from the
cross-over.

In addition to the above subtlety, the procedure of taking $V\rightarrow\infty$
limit\cite{FK3} may cause some systematic uncertainties 
in the estimation of the critical endpoint. 
For example, an extrapolation linear in $1/V$
gives 
Im $\beta_0^\infty = 0.0018(7)$ at $\mu a = 0.16$,
suggesting that this point corresponds to the cross-over. 
As shown in the right of Fig.~\ref{fig:compare}, however,
the quadratic fit or the linear fit for larger 3 volumes 
gives Im $\beta_c^\infty = -0.0004(11)$ or $0.0003(9)$, respectively,
concluding that the phase transition is consistent with the first order.
Clearly
a more careful finite volume analysis is required for a definite conclusion
on the location of the critical endpoint.


\section{Conclusions}
By recent developments for lattice QCD techniques,
numerical simulations become possible at small $\mu$ and non-zero $T$.
Preliminary  results suggest $\mu_B^E \simeq$ 350 -- 450 MeV at physical 
light and strange quark masses. However further confirmations will be 
definitely required for the reliable estimate of $\mu_B^E$.
So far numerical simulations at non-zero $\mu$ have been performed
only with KS fermions. Therefore new calculations by other fermion 
formulations such as Wilson/clover or domain-wall/overlap fermions
will be necessary to check the present results by KS fermions.
Definitely new ideas will be needed to explore QCD phase structure
at large $\mu$ and low $T$.


\section*{Acknowledgments}
This work is supported in part by the Grant-in-Aid of the Ministry of
Education(Nos. 13135204, 15540251, 16028201).









\begin{thebibliography}{0}



\bibitem{hands} S.~Hands, {\it Nucl. Phys.} {\bf B (Proc. Suppl.)106\& 107},
142 (2002).

\bibitem{ejiri1} S.~Ejiri, {\it Nucl. Phys.} {\bf B (Proc. Suppl.)94},
19 (2001).

\bibitem{glasgow} I.M.~Barbour and A.J.~Bell, {\it Nucl. Phys/}
{\bf B372} 385 (1992); 
I.M.~Barbour {\it et al.},  {\it Nucl. Phys/}
{\bf B(Proc. Suppl.) 60A} 220 (1998). 

\bibitem{GG}R.~V.~Gavai and S.~Gupta, {\it Phys. Rev.} {\bf D68}, 034506,
(2003); {\bf D71}, 114014 (2005).

\bibitem{FK1}Z.~Fodor and S.D.~Katz, {\it Phys. Lett.} {\bf B534}, 87 (2002).
\bibitem{FK2}Z.~Fodor and S.D.~Katz, {\it JHEP}{\bf 0203}, 014 (2002).
\bibitem{FK3}Z.~Fodor and S.D.~Katz, {\it JHEP}{\bf 0404}, 050 (2004).

\bibitem{B-S}C.R.~Alton {\it et al.}, {\it Phys. Rev.}
{\bf D66}, 074507 (2002).

\bibitem{DEL}M.~D'Elia and M-P.~Lombardo, {\it Phys. Rev.}
{\bf D67}, 014505 (2003).

\bibitem{FP} P.~de~Forcrand and O.~Philipsen, {\it Nucl. Phys.}
{\bf B642}, 290 (2002).

\bibitem{LY} C.~N.~Yang and T.~D.~Lee, {\it Phys. Rev.}
{\bf 87}, 404 (1952).

\bibitem{B-S2}S.~Ejiri {\it et al.}, {\it Prog. Theor. Phys. Suppl.}
{\bf 153}, 118 (2004); F.~Karsch {\it et al.}, {\it Nucl. Phys.}
{\bf B(Proc. Suppl.)129}, 614 (2004).

\bibitem{FP2}  P.~de~Forcrand and O.~Philipsen, {\it Nucl. Phys.}
{\bf B673}, 170 (2003).

\bibitem{ejiri2} S.~Ejiri, {\it Phys. Rev.} {\bf D69}, 094506 (2004).

\end{thebibliography}
\end{document}